\documentclass[12pt,twoside]{article}
\usepackage{amsmath,mathrsfs,bm,amssymb,color}
\newtheorem{theorem}{Theorem}[section]
\newtheorem{proposition}[theorem]{Proposition}
\newtheorem{lemma}[theorem]{Lemma}
\newtheorem{corollary}[theorem]{Corollary}
\newtheorem{definition}[theorem]{Definition}
\newtheorem{example}[theorem]{Example}
\newtheorem{remark}[theorem]{Remark}

\newtheorem{assumption}[theorem]{Assumption}
\newcommand{\BERN}{\ms B_0}
\newcommand{\eq}[1]{\begin{equation}\label{#1}}
\newcommand{\en}{\end{equation}}
\newcommand{\eqn}{\begin{eqnarray*}}
\newcommand{\enn}{\end{eqnarray*}}
\newcommand{\eqnn}{\begin{eqnarray}}
\newcommand{\ennn}{\end{eqnarray}}
\newcommand{\proof}[1][Proof]{{\sc #1.} }
\newcommand{\qed}{\hfill {\bf qed}\par
\medskip}
\newcommand{\bi}{\begin{description}}
\newcommand{\ei}{\end{description} }
\newcommand{\bern}{\Psi}

\def\bbbone{{\mathchoice {\rm 1\mskip-4mu l} {\rm 1\mskip-4mu l}
{\rm 1\mskip-4.5mu l} {\rm 1\mskip-5mu l}}}
\def\one{\bbbone}

\newcommand{\lkkk }{\left[}
\newcommand{\rkkk}{\right]}

\newcommand{\BM}{\Omega_P}

\newcommand{\SU}{\Omega_\nu}
\newcommand{\dx}{{\rm d}x}
\newcommand{\dy }{{\rm d}y}
\newcommand{\ds}{{\rm d}s}
\newcommand{\dt }{{\rm d}t}
\newcommand{\du}{{\rm d}u}
\newcommand{\dr}{{\rm d}r}
\newcommand{\dxi }{{\rm d}\xi}

\newcommand{\ix}{\int_{\BR}\!\!\!\dx }

\newcommand{\bl}[1]{\begin{lemma}\label{#1}}
\newcommand{\el}{\end{lemma}}
\newcommand{\bc}[1]{\begin{corollary}\label{#1}}
\newcommand{\ec}{\end{corollary}}
\newcommand{\bt}[1]{\begin{theorem}\label{#1}}
\newcommand{\et}{\end{theorem}}
\newcommand{\bp}[1]{\begin{proposition}\label{#1}}
\newcommand{\ep}{\end{proposition}}
\newcommand{\br}[1]{\begin{remark}\label{#1}}
\newcommand{\er}{\end{remark}}
\newcommand{\bd}[1]{\begin{definition}\label{\rm #1}}
\newcommand{\ed}{\end{definition}}

\newcommand{\DEL}{\Delta}
\newcommand{\AP}{\alpha}
\newcommand{\GA}{\Gamma}

\newcommand{\E}{\mathbb E}

\newcommand{\RR}{{\mathbb  R}}
\newcommand{\BR}{{{\mathbb  R}^d }}

\newcommand{\kak}[1]{(\ref{#1})}
\newcommand{\LR}{{L^2(\BR)}}

\newcommand{\lk}{\left(}
\newcommand{\rk}{\right)}
\newcommand{\lkk}{\left\{}
\newcommand{\rkk}{\right\}}
\newcommand{\ms}[1]{\mathscr{#1}}
\newcommand{\la}{\lambda }
\newcommand{\EE}{\mathbb E}
\newcommand{\ov}[1]{\overline{#1}}

\newcommand{\han}{{1/2}}

\newcommand{\Tr}{\mathop{\mathrm{Tr}}\nolimits}
\newcommand{\Spec}{\mathop{\mathrm{Spec}}\nolimits}
\newcommand{\Ker}{\mathop{\mathrm{Ker}}\nolimits}

\newcommand{\pro}[1]{(#1_t)_{t\geq0}}

\newcommand{\s}{\sigma}

\renewcommand{\d}{\displaystyle}
\newcommand{\non}{\nonumber}

\makeatletter
\@addtoreset{equation}{section}
\makeatother

\title
{\Large \sc Lieb-Thirring Bound for Schr\"odinger Operators with
Bernstein Functions of the Laplacian}
\author{
\small Fumio Hiroshima\\
{\small\it Faculty of Mathematics, Kyushu University}    \\[-0.7ex]
{\small\it  744 Motooka Fukuoka, 819-0395,  Japan}      \\[-0.7ex]
{\small  {\tt hiroshima@math.kyushu-u.ac.jp}}\\[0.3cm]
\small J\'ozsef L\H{o}rinczi\\
{\it \small School of Mathematics, Loughborough University} \\[-0.7ex]
{\it \small Loughborough LE11 3TU, United Kingdom} \\[-0.7ex]
{\small {\tt  J.Lorinczi@lboro.ac.uk}} \\[-0.7ex]}

\date{}

\begin{document}
\maketitle
\setlength{\baselineskip}{14pt}

\bigskip\bigskip\bigskip
\begin{abstract}
\noindent
A Lieb-Thirring bound for Schr\"odinger operators with Bernstein functions of the Laplacian is shown
by functional integration techniques. Several specific cases are discussed in detail.

\end{abstract}

\newpage
\section{Introduction}
In mathematical physics there is much interest in an inequality due originally to Lieb and Thirring giving
an upper bound on the number of bound states for a Schr\"odinger operator $-\frac{1}{2}\Delta + V$. With
$N_0$ denoting the number of non-positive eigenvalues of the Schr\"odinger operator, in a semi-classical
description it is expected that
\begin{align}
N_{0}(V)
=
\frac{1}{(2\pi)^d}\int_{\BR\times \BR}\one_{\{(p,x): \; |p|^2+V(x)\leq 0\}}
{\rm d}p\dx.
\end{align}
The right hand side above is computed as
\begin{align}
\frac{1}{(2\pi)^d}\int_{\BR} \dx \int _{\BR} \one_{\left\{|\xi|\leq \sqrt{V_-(x)}\right\}}
{\rm d}\xi= \frac{1}{(2\pi)^d}\frac{\s(S_{d-1})}{d} \int_\BR |V_-(x)|^{d/2} \dx
\end{align}
where $\s\!\lk {\rm S}_{d-1}\rk = \frac{2\pi^{d/2}}{\GA(d/2)}$ and $V_-$ is the negative part of $V$. The
Lieb-Thirring inequality then says that
\begin{align}
N_0(V) \leq {C_d}\int_\BR |V_-(x)|^{d/2}\dx,
\end{align}
see \cite{lie76,lie80}, where $C_d$ is a constant dependent on $d$ alone. Various extensions have been further
studied by many authors, see \cite{LS10} and references therein.

Following our work \cite{hil09} in which we defined generalized Schr\"odinger operators of the form
\eq{li}
H = \bern\left(-\frac{1}{2}\Delta\right)+V
\en
where $\bern$ denotes a Bernstein function (see below), it is a natural question if a similar Lieb-Thirring
bound can be established and how does this depend on the choice of the Bernstein function. We will actually
derive under some conditions that
\begin{align}
N_0(V)\leq A\int_\BR\lk \Psi^{-1}(|V(x)|)\rk^{d/2}\dx
\end{align}
(Theorem \ref{main18-1} and Corollary \ref{cor1}) by using estimates of the diagonal part of the heat kernel of
subordinate Brownian motion generated by $\bern\left(-\frac{1}{2}\Delta\right)$. This extension includes beside
usual Schr\"odinger operators also fractional Schr\"odinger operators of the form $(-\Delta)^{\alpha/2}+V$ and
relativistic Schr\"odinger operators $(-\Delta+m^2)^{1/2}-m +V$. General Bernstein functions receive increasing
attention in the study of stochastic processes with jump discontinuities and their potential theory \cite{SSV10}.

A Lieb-Thirring bound for generalized kinetic energy terms was first obtained in \cite{dau83}. Although the author
mentions that similar bounds can be derived for generalizations using \kak{li}, the focus of that paper is primarily
the relativistic Schr\"odinger operators above with or without mass. Lieb-Thirring inequalities for fractional
Schr\"odinger operators compensated by the Hardy weight have been obtained more recently in \cite{FLS08} by using
methods of Sobolev inequalities. A reference considering the same problem for relativistic Schr\"odinger operators
including magnetic fields is \cite{imp}.

The remainder of this paper is organized as follows.
In Section 2 we recall
the definition of such Schr\"odinger operators and briefly describe the stochastic processes related to them.
In the main Section 3 we state and prove the Lieb-Thirring inequality for this class of operators, and obtain
some explicit variants. In Section 4 we discuss some cases of special interest.

\section{Schr\"odinger operators with Bernstein functions of the Laplacian}
Consider the function space
$$
\ms B = \left\{ \Psi\in C^\infty(\RR^+): \; \Psi(x) \geq 0, \; (-1)^n
\left( \frac{{\rm d}^n\Psi}{\dx^n}\right) (x) \leq 0,   \forall n=1,2,...\right\}
$$
An element of $\ms B$ is called a \emph{Bernstein function}. We also define the subclass $\BERN =
\lkk f\in \ms B: \, \lim_{u\rightarrow 0+}f(u)=0\rkk$.

Bernstein functions in $\ms B_0$ have the following integral representation. Let $\ms L$ be the set of Borel
measures $\la$ on $\RR\setminus\{0\}$ such that $\la((-\infty,0))=0$ and $\d \int_{\RR\setminus\{0\}}
(y\wedge1)\la (\dy )<\infty$. Note that every $\la\in \ms L$ is a L\'evy measure. Then it can be shown that
for every Bernstein function $\bern\in \BERN $ there exists $(b,\la)\in [0,\infty) \times \ms L$ such that
\eq{6}
\bern(u)=bu+\int_0^\infty (1-e^{-uy})\la(\dy ).
\en
Conversely, the right hand side of \kak{6} is in $\BERN $ for each pair $(b,\la)\in [0,\infty)\times \ms L$.
It is known that the map $\BERN \rightarrow [0,\infty)\times\ms L$, $\bern\mapsto(b,\la)$ is bijective.

Next consider a probability space $(\SU, \ms F_\nu, \nu)$ and a stochastic process $(T_t)_{t\geq 0}$ on it.
Recall that $(T_t)_{t\geq 0}$ is called a subordinator whenever it is a L\'evy process starting at 0, and
$t\mapsto T_t$ is almost surely a non-decreasing function. Let $\ms S$ denote the set of subordinators on
$(\SU,\ms F_\nu,\nu)$. Also, let $\bern\in \BERN $ or, equivalently, a pair $(b,\la)\in [0,\infty)\times
\ms L$ be given. Then by the above bijection there is a unique $(T_t )_{t\geq 0}\in \ms S$ such that
\eq{10}
\EE_{\nu}^0[e^{-u T_t}]=e^{-t \bern(u)}.
\en
Conversely, for every $(T_t)_{t\geq0}\in \ms S$ there exists a unique $\bern\in \BERN $, i.e., a pair $(b,\la)\in
[0,\infty)\times \ms L$ such that \kak{10} is satisfied. In particular, (\ref{6}) coincides with the
L\'evy-Khintchine formula for Laplace exponents of subordinators.
Using the bijection between $\BERN $ and $\ms S$, we denote by $T_t^\bern$ the subordinator uniquely associated
with $\bern \in \BERN $.

It is known that the composition of a Brownian motion and a subordinator yields  a L\'evy process. This process is
$
X_t:\BM\times \SU \ni (\omega_1,\omega_2) \mapsto B_{T _t(\omega_2)}(\omega_1)\in
\RR^d
$
called $d$-dimensional subordinate Brownian motion with respect to the subordinator $(T_t)_{t\geq 0}$. Its properties
are determined by
$
\EE_{P\times \nu}^{0}[e^{i\xi\cdot X_t}]=e^{-t\bern(|\xi|^2/2)}$.
The function
\begin{equation}
{\rm P}^\bern_t(x)=\frac{1}{(2\pi)^d} \int_\BR e^{-ix\cdot \xi}e^{-t\bern(|\xi|^2/2)}\dxi
\label{pt}
\end{equation}
gives the distribution of $X_t$ in $\RR^d$.

Let $h=-\Delta$ be the Laplacian in $\LR$. We assume throughout this paper that $d\geq 3$.
 Define the operator $\bern(h/2)$ on $\LR$ with Bernstein
function $\bern \in \BERN$. Let $V=V_+-V_-$, where $V_+=\max\{V,0\}$, $V_-=\min\{-V,0\}$, and assume that $V_-$ is
form-bounded with respect to $\bern(h/2) $ with a relative bound strictly smaller than 1, and $V_+\in L_{\rm loc}^1(\BR)$.
Then we define the \emph{Schr\"odinger operator with Bernstein function $\bern$ of the Laplacian} by
\eq{p1}
H^\bern=\bern(h/2)  \,\,\dot +\, \,   V_+ \, \, \dot -\,\,   V_-.
\en
In what follows we simply write $H^\bern=\bern(h/2)+V$ instead of
\kak{p1}.
\bp{f}
We have the functional integral representation for the semigroup $e^{-tH^\bern}$, $t\geq 0$, given by
\eq{bo81}
(f, e^{-tH^\bern   }g)= \ix \mathbb E^{x}_{P\times\nu} \lkkk \ov{f(X_0)}g(X_t)
e^{-\int_0^t V(X_s)\ds } \rkkk.
\en
\ep
\proof
This is obtained by subordination and an application of the Trotter product formula combined with a
limiting argument. For a detailed proof we refer to \cite{hil09,LHB09}.
\qed


In view of applications (quantum theory, anomalous transport theory, financial mathematics etc) some
particular choices of Bernstein functions are of special interest involving the following stochastic
processes:
\begin{enumerate}
\item[(1)]
\emph{symmetric $\alpha$-stable processes:} $\Psi(u) = (2u)^{\alpha/2}$, $0 < \alpha \leq 2$

\item[(2)]
\emph{relativistic $\alpha$-stable processes:} $\Psi(u) = (2u + m^{2/\alpha})^{\alpha/2}$, with $m > 0$


\item[(3)]
\emph{jump-diffusion processes:} $\Psi(u) = au  + b u ^{\alpha/2}$, with $a,b \in \RR$.
\end{enumerate}

\section{Lieb-Thirring  bound}
The following is a standing assumption throughout the paper.
\begin{assumption}
\label{v}
\hspace{0cm}
\bi
\item[(1)]
$V$ is a continuous and non-positive function
\item[(2)]
there exists $\lambda^\ast>0$ such that $\|(\bern(h/2)+\lambda)^{-\han}
|V|^\han\|<1$ for all $\lambda \geq \lambda^\ast$
\item[(3)]
the operator $(\bern(h/2)+\lambda)^{-\han} |V|^\han$ is compact for all $\lambda \geq0$
\item[(4)]
there exists $n_0>0$ such that $\Tr (|V|^\han (\bern(h/2)+\lambda)^{-1}|V|^\han)^n<\infty$
for all $n\geq n_0$ and $\lambda>0$.
\ei
\end{assumption}
Part (2) of Assumption \ref{v} implies that $V$ is relatively form bounded with respect to $\bern(h/2)$
with relative bound strictly smaller than 1. Part (3) ensures that the Birman-Schwinger principle \kak{bs2}
holds, and (4) is used in the proof of Lemma \ref{generic}.
\begin{example}
\rm{
Let $L^{\infty,0}(\BR)$ be the set of functions $f\in L^\infty(\BR)$ such that $\d \lim_{|x|\to\infty} |f(x)|=0$.
It is well known that if $P,Q\in L^{\infty,0}(\BR)$, then $P(-i\nabla) Q(x)$ is a compact operator \cite{sim04}.
Thus $(\bern(h/2)+\lambda)^{-\han} |V|^\han$ is compact for $V\in L^{\infty,0}$, since $\bern$ is increasing.
Moreover, if $\bern(h/2)=-\Delta$ and $V\in L^{d/2}(\BR)$, (4) of Assumption \ref{v} is satisfied with $n_0=d/2$.
}
\end{example}

Consider the number
\eq{p2}
N_E(V)={\rm dim}\one_{(-\infty,-E]}(H^\bern).
\en
In the original context of quantum theory this expression has the relevance of counting the number of
bound states of energy up to $-E<0$. Recall \cite{sim05} that the Birman-Schwinger kernel is defined by
\eq{bs}
K_E=|V|^{\han}(\bern(h/2)+E)^{-1}|V|^\han
\en
and the Birman-Schwinger principle says that
\eq{bs2}
\begin{array}{ll}
\d N_E(V)={\rm dim}\one_{[1,\infty)}(K_E),&-E<0\\ \\
\d N_0(V) \leq {\rm dim}\one_{[1,\infty)}(K_0),&E=0.
\end{array}
\en
\begin{example}
\rm{
Let $V=V_+-V_-$ be such that $V_-\in L^\infty(\BR)$.
Since
$\bern(h/2)-V_-\leq H^\bern$,
the number of negative eigenvalues of $H^\bern$ is smaller than that of
$H^\bern_-=\bern(h/2)-V_-$.
So instead of $H^\bern$, we consider $H_-^\bern$.
 Since $|V_-|\in L^\infty$,
$(\bern(h/2)+\lambda)^{-\han} |V_-|^\han$  is compact.
Thus the Birman-Schwinger principle can be applied to $H_-^\bern$.
}
\end{example}

Let $F_\lambda(x)=x(1+\la x)^{-1}=x\int_0^\infty e^{-y(1+\la x)} \dy $ and $g_\la(x)=e^{-\la x}$. The two
functions are related by
\eq{rel}
F_\la(x)=x\int_0^\infty e^{-y} g_\la(xy) \dy .
\en
By a direct computation we obtain
\eq{bs3}
F_\la(K_E)=|V|^\han (\bern(h/2)+\la |V|+E)^{-1}|V|^\han
\en
and by Laplace transform
\eq{bs4}
\lk
F_\la(K_E)u \rk (x) = |V(x)|^\han \lk \int_0^\infty  \dt  e^{-tE} e^{-t(\bern(h/2)+\la |V|)}
|V|^\han u \rk (x)
\en
follows. By \kak{bs2} we have
\begin{eqnarray*}
N_E(V)&=&\#\{F_\la(\mu)|F_\la(\mu) \mbox{ is an eigenvalue of } F_\la(K_E) \mbox{ and }\mu\geq 1\},\quad E>0\\
N_0(V)&\leq&\#\{F_\la(\mu)|F_\la(\mu) \mbox{ is an eigenvalue of } F_\la(K_0) \mbox{ and }\mu\geq 1\},\quad E=0.
\end{eqnarray*}
Since $F_\la$ is monotone increasing, it follows that
\eq{bs16}
N_E(V) \leq \frac{1}{F_\la(1)}\sum_{\mu\in \Spec(K_E) \atop \mu\geq 1} F_\la(\mu).
\en
Using this we will estimate the trace of $F_\la(K_E)$. From Theorem \ref{f} we obtain
\eq{bs5}
\lk
F_\la(K_E)u \rk (x)= |V(x)|^\han \int_0^\infty  \dt  e^{-tE}
\E_{P\times\nu}^x\left[e^{-\la\int_0^t |V(X_s)| \ds } |V (X_t)|^\han u(X_t) \right].
\en
In order to express the kernel of $e^{-t(\bern(h/2)+\lambda |V|)}$ in terms of a conditional expectation we
use the following notation.  Let $\E_{P\times \nu}^0[Y |X_t]$ be conditional expectation with respect to the
$\s$-field $\s(X_t)$, i.e., $\E_{P\times \nu}^0[Y |X_t]$ is measurable with respect to $\s(X_t)$. Generally,
a function $f$ measurable with respect to $\s(X_t)$ can be written as $f=g(X_t)$ with a suitable function $g$.
We write $\E_{P\times \nu}^0[Y |X_t]=g(X_t)$, and use the notation $g(x) = \E_{P\times \nu}^0[Y |X_t=x]$, i.e.,
$\E_{P\times \nu}^0[Y |X_t]=\int \E_{P\times \nu}^0[Y |X_t=x] P_t^\bern(x) dx$. In these terms we then have
\eq{bs19}
e^{-t(\bern(h/2)+\lambda |V|)}(x,y)=\E_{P\times\nu}^0\left[\left.
e^{-\la\int_0^t |V(X_s+x)| \ds }\right|X_t+x=y\right]{\rm P}^\bern_t(x-y),
\en
where ${\rm P}^\bern_t$ is the distribution of $X_t$ given by \kak{pt}.

\bl{hiroshima2}
The map $ (x,y)\mapsto e^{-t(\bern(h/2)+|V|)}(x,y)$  is continuous.
\el
\proof
Let $P_{[0,T]}^{x,y}$ denote Brownian bridge measure starting from $x$ at $t=0$ and
ending in $y$ at $t=T$. Then by the Feynman-Kac-like formula \kak{bo81} and using that $X_s = B_{T_s}$
we see that
\eq{h1}
(f, e^{-t(\bern(h/2)+|V|)}g)=\int_{\RR^d \times \RR^d} \!\!\!  \bar f(x) g(y)
\E_\nu\left[\Pi_{T_t}(x-y)\E_{P_{[0,T_t]}^{x,y}}[e^{-\int_0^t|V(B_{T_s})| \ds }]\right]\dx  \dy ,
\en
where $\Pi_{t}(x)$ is the Gaussian heat kernel. Note that the measure $P_{[0,T_t]}^{x,y}=
P_{[0,T_t(\omega_2)]}^{x,y}$ is defined for every $\omega_2\in \Omega_\nu$. For every $\omega_2
\in\Omega_\nu$ we also define the Brownian bridge $\pro Z$ by
$$
Z_t=\lk 1-\frac{t}{T_t}\rk x+\frac{t}{T_t}y -\frac{t}{T_t}B_{T_t}+B_t,
$$
where $T_t$ depends  on $\omega_2$. Thus \kak{h1} is equal to
\eq{h2}
(f, e^{-t(\bern(h/2)+|V|)}g)=\int_{\RR^d \times \RR^d}\!\!\!  \bar f(x) g(y) \E_\nu\left[\Pi_{T_t}(x-y)
\E_P^0 [e^{-\int_0^t|V(Z_s)| \ds }]\right] \dx  \dy.
\en
Hence the integral kernel is given by
$$
e^{-t(\bern(h/2)+|V|)}(x,y)=\E_\nu\left[\Pi_{T_t}(x-y)\E_P^0[e^{-\int_0^t|V(Z_s)| \ds }]\right]
$$
and implies joint continuity with respect to $(x,y)$.
\qed
From Lemma \ref{hiroshima2} it follows that the kernel of $F_\la(K_E)$,
\begin{eqnarray}
\lefteqn{
F_\la(K_E)(x,y) = |V(x)|^\han|V(y)|^\han\non }
\\
\label{bs6}
&& \times \int_0^\infty \dt  e^{-tE}
\E_{P\times\nu}^0\left[\left. g_\la\lk \int_0^t|V(X_s+x)|\ds \rk \right|X_t+x=y \right]{\rm P}^\bern_t(x-y)
\end{eqnarray}
is also jointly continuous in $(x,y)$. Here we used that $ g(x)=e^{-\la x}$. By setting $x=y$ in \kak{bs6}
it is seen that $\Tr  F_\la(K_E)=\int_\BR F_\la(K_E)(x,x)\dx$. This gives the expression
\eq{bs}
\Tr F_\la(K_E)=\int_{\RR^d} \dx  |V(x)|\int_0^\infty \dt  e^{-tE}
\E_{P\times\nu}^0\left[\left.g_\la\lk \int_0^t|V(X_s+x)|\ds \rk\right|X_t=0 \right]{\rm P}^\bern_t(0).
\en
\bl{1}
It follows that
\eq{bs10}
\Tr F_\la(K_E)=\int_{\RR^d} \dx   \int_0^\infty \frac{dt }{t} {e^{-tE}}
\E_{P\times\nu}^0 \left[\left. G_\la \lk \int_0^t|V(X_s+x)|\ds  \rk \right|X_t=0 \right] {\rm P}^\bern_t(0),
\en
where $G_\la(x)=xg_\la(x)=xe^{-\la x}$.
\el
\proof
It suffices to show that
\begin{eqnarray}
\lefteqn{
\hspace{-2cm}
\frac{1}{t} \int_{\RR^d} \dx  \E_{P\times\nu}^0\left[ \left.e^{-\int_0^t|V(X_s+x)|\ds }\int_0^t |V(X_r+x)|\dr \right|
X_t=0 \right]{\rm P}^\bern_t(0) } \non \\
&&\label{bs17}
= \int_{\RR^d} \dx  |V(x)| \E_{P\times\nu}^0\left[\left. e^{-\int_0^t|V(X_s+x)|\ds } \right| X_t=0\right]
{\rm P}^\bern_t(0).
\end{eqnarray}
Let $U_r=e^{-r(\bern(h/2)+|V|)}|V| e^{-(t-r)(\bern(h/2)+|V|)}$ for $0\leq r\leq t$. Note that $U_r$ is
compact and thus $\Tr  U_r = \Tr U_0$. By the Markov property of $(X_t)_{t\geq 0}$ it follows that
\begin{eqnarray*}
\lk U_rf\rk (x)
&=&
\E_{P\times\nu}^x\left[e^{-\int_0^r |V(X_s)| \ds } |V(X_r)|
\E_{P\times\nu}^{X_r}\left[e^{-\int_0^{t-r}|V(X_s)| \ds }f(X_{t-r})\right]\right]\\
&=&
\E_{P\times\nu}^x\left[e^{-\int_0^t |V(X_s)| \ds } |V(X_r)|f(X_t)
\right].
\end{eqnarray*}
Thus the right hand side above is expressed as
$$
=\int_{\RR^d} {\rm P}^\bern_t (x-y)\E_{P\times\nu}^0\left[\left. e^{-\int_0^t|V(X_s+x)|\ds } |
V(X_r+x)|\right|X_t+x=y\right]f(y) \dy.
$$
This furthermore gives
\eq{bs8}
\Tr U_r=\int_0^t \frac{\dr }{t} \Tr  U_r= \frac{1}{t}\int_{\RR^d} \dx  {\rm P}^\bern_t(0)
\E_{P\times\nu}^0\left[\left. e^{-\int_0^t|V(X_s+x)|\ds } \int_0^t |V(X_r+x)|\dr  \right| X_t=0\right],
\en
where we interchanged $\dr $ and ${\rm d}P^0$. Equality $\int_0^t \frac{\dr }{t} \Tr  U_r= \Tr U_0 $
together with  \kak{bs8} yield \kak{bs17}. Hence the lemma follows.
\qed
We may vary $F_\la$ and $g_\la$ while keeping relationship \kak{rel} unchanged. Let $F:[0,\infty)\to [0,\infty)$
be a strictly increasing function such that
\eq{rel2}
F(x)=x\int_0^\infty e^{-y}g(xy)\dy ,
\en
where $g$ is a non-negative function on $\RR$. Write
\eq{rel3}
G(x)=x g(x).
\en
\bl{l2}
Let Assumption \ref{v} hold and take any $F$, $G$ and $g$ satisfying \kak{rel2}. Suppose that $G$ is non-negative
and lower semi-continuous. Then it follows that
\eq{bs11}
\Tr F(K_E)= \int_{\RR^d} \dx   \int_0^\infty \dt  \frac{e^{-tE}}{t}
\E_{P\times\nu}^0\left[\left. G \lk \int_0^t|V(X_s+x)|\ds  \rk \right|X_t=0 \right] {\rm P}^\bern_t(0).
\en
\label{generic}
\el
\proof
The proof is obtained by a slight modification of \cite[Theorem 8.2]{sim04} and \cite[Lemma 3.51]{LHB09}.
\qed
\bt{main}
{\bf (Lieb-Thirring bound)}
Let Assumption \ref{v} hold, $F$, $G$ be any functions satisfying \kak{rel2} and \kak{rel3}, and $G$ furthermore
be convex. Then
\eq{main18}
N_0(V) \leq \frac{1}{F(1)} \int_0^\infty \frac{\ds }{s} G(s) \int_{\RR^d} {\rm P}^\bern_{s/|V(x)|}(0)\one_{\{|V(x)|>0\}} \dx,
\en
where
$$
{\rm P}^\bern_{s/|V(x)|}(0)=(2\pi)^{-d}\int_\BR e^{-s\bern(|\xi|^2/2)/|V(x)|}\dxi.
$$
\et
We note that the right hand side of \kak{main18} may not be finite, this depends on the choice of the convex function $G$.

\medskip
\noindent
\proof
Since $F$ is a monotone increasing function, we have
\begin{eqnarray*}
N_0(V)
&\leq& \frac{1}{F(1)} \Tr (F(K_0))\\
&=&
\frac{1}{F(1)} \int_0^\infty \frac{\dt }{t} \int_{\RR^d} \dx
\E_{P\times\nu}^0\left[\left. G \lk \int _0^t t |V(X_s+x)|\frac{\ds }{t} \rk \right|X_t=0 \right]{\rm P}^\bern_t(0).
\end{eqnarray*}
 Then by the Jensen inequality
$$
N_0(V)\leq \frac{1}{F(1)} \int_0^\infty \frac{\dt }{t} \int_{\RR^d} \dx  \E_{P\times\nu}^0\left[\left.\int_0^t \frac{\ds }{t}
G\lk t|V(X_s+x)|\rk \right|X_t=0 \right]{\rm P}^\bern_t(0).
$$
Using that $\int_0^t \frac{\ds }{t}=1$ and swapping $\dx $ and $dP^0\times d\nu$, we obtain
\begin{eqnarray*}
N_0(V)
&\leq&
\frac{1}{F(1)} \int_0^\infty {\rm P}^\bern_t(0)\frac{\dt}{t} \int_{\RR^d} G(t|V(x)|)\dx .
\end{eqnarray*}
When $V(x)=0$, also $G (tV(x))=0$. This implies that the right hand side above equals
$$
\frac{1}{F(1)} \int_0^\infty {\rm P}^\bern_t(0)\frac{\dt}{t} \int_{\RR^d} G(t|V(x)|)\one_{\{|V(x)|>0\}}\dx .
$$
Changing the variable from $t|V(x)|$ to  $s$ and integrating with respect to $s$, we obtain
\kak{main18}.
\qed

Next we are interested to see how the Lieb-Thirring bound (\ref{main18}) in fact depends on the Bernstein
function $\Psi$. To make this expression more explicit we note that the diagonal part of the heat kernel
has the representation \cite{JKLS}
\begin{align}\label{jacob}
P_t^\bern(0)=(2\pi)^{-d} \int_0^\infty  e^{-r}\lk
\int_{\RR^d} \one_{\left\{\sqrt{{\bern(\xi^2/2)}}\leq \sqrt{r/t}\right\}}\dxi\rk \dr.
\end{align}
Denote by $\Bbb B^\Psi(x,r)$ a ball of radius $r$ centered in $x$ in the topology of the metric
$$
d^\bern(\xi,\eta) =\sqrt{\bern(|\eta-\xi|^2/2)}.
$$
Notice that $d^\bern(\xi,\eta)=0$ if and only if $\xi=\eta$, since $\bern$ is concave and a
$C^\infty$-function. Then the integral $\int_{\RR^d} \one_{\left\{\sqrt{{\bern(\xi^2/2)}} \leq
\sqrt{r/t}\right\}}\dxi$ is the volume of $\Bbb B^\bern (0,\sqrt{r/t})$ in this metric. If $d^\bern$ satisfies
the condition
$$
\int_{\RR^d} \one_{\Bbb B^\bern(x,2r)}\dy \leq c \int_{\RR^d} \one_{\Bbb B^\bern(x,r)} \dy, \quad x \in \RR^d,
  r > 0
$$
with a constant $c > 0$ independent of $x$ and $r$, then $d^\bern$ is said to have the volume doubling property.
When $d^\bern$ has this property, then furthermore it follows that
\begin{equation}
\label{volumedoubling}
c_1\int_{\RR^d} \one_{\left\{\sqrt{{\bern(\xi^2/2)}}\leq \sqrt{r/t}\right\}} \dxi
\leq P_t^\bern(0) \leq
c_2\int_{\RR^d} \one_{\left\{\sqrt{{\bern(\xi^2/2)}}\leq \sqrt{r/t}\right\}}  \dxi
\end{equation}
with some constants $c_1$ and $c_2$. A necessary and sufficient condition for $\Psi \in \BERN$ to give rise to a
volume doubling $d^\bern$ is
$$
\liminf_{u\to 0} \frac{\Psi(Cu)}{\Psi(u)} > 1 \quad \mbox{and} \quad
\liminf_{u\to \infty} \frac{\Psi(Cu)}{\Psi(u)} > 1
$$
for some $C>1$. In particular, this implies that $\Psi$ increases at infinity as a (possibly fractional) power.
For details, we refer to \cite{JKLS}.
\bt{main2}
Suppose that $\bern\in\ms B_0$ is strictly monotone increasing. Then under the assumptions of Theorem \ref{main}
we have
\eq{main18-1}
N_0(V) \leq \frac{2^{\frac{3d}{2}+1} \pi^{\frac{d}{2}}}{d\Gamma(\frac{d}{2}) F(1)}
\int_0^\infty \frac{\ds }{s} G(s) \int_{\RR^d} \dx
\int _0^\infty \left(\Psi^{-1}\left(\frac{r|V(x)|}{s}\right)\right)^{d/2} e^{-r} \dr.
\en
Furthermore, if $d^\bern$ has the volume doubling property, then
\eq{main18-2}
N_0(V) \leq c_2 \frac{2^{\frac{3d}{2}+1} \pi^{\frac{d}{2}}}{d\Gamma(\frac{d}{2}) F(1)}
\int_0^\infty \frac{\ds }{s} G(s) \int_{\RR^d}
\left(\Psi^{-1}\left(\frac{|V(x)|}{s}\right)\right)^{d/2}\dx.
\en
\et
\proof
Since under the assumption the function $\Psi \in \BERN$ is invertible and its inverse is increasing, the
proof is straightforward using $\Ker\bern=\{0\}$, (\ref{jacob}) and (\ref{volumedoubling}).
\qed

In the case when $\bern\in\ms B_0$ has a scaling property, we can derive a more explicit formula.
\bc{cor1}
Suppose that $\bern\in\ms B_0$ is strictly monotone increasing and  the assumptions of Theorem \ref{main}
hold. In addition, assume that there exists $\gamma > 0$ such that $\bern (a u)=a^\gamma \bern(u)$ for all
$a, u \geq 0$. Then
\eq{main18-4}
N_0(V) \leq A \int_{\RR^d} \left(\Psi^{-1}\left(|V(x)|\right)\rk^{d/2} \dx,
\en
where $\d A=
\frac{2^{\frac{3d}{2}+1}\pi^{\frac{d}{2}}\Gamma(\frac{d}{2\gamma}+1)}{d\Gamma(\frac{d}{2}) F(1)}
\int_0^\infty G(s)s^{-1-\frac{d}{2\gamma}} \ds$.
\label{scale}
\ec
\proof
The inverse function $\bern^{-1}$ has the scaling property $\bern^{-1}(av)=a^{1/\gamma}\bern^{-1}(v)$.
Thus the corollary follows.
\qed
Instead of the scaling property suppose now that there exists $\lambda > 0$ such that $\Psi(u) \geq
C u^\lambda$ with a constant $C>0$. This inequality holds for at least large enough $u$ if $d^\bern$
has the volume doubling property. Then we have a similar formula to that in Corollary \ref{cor1}.
\bc{cor2}
Suppose that $\bern\in\ms B_0$ is strictly monotone increasing and the assumptions of Theorem \ref{main}
hold. If $\Psi(u) \geq C u^\lambda$, then
\begin{align}
N_0(V) \leq A \int_{\RR^d} |V(x)|^{d/2\lambda}\dx,
\end{align}
where $\d A =  \frac{2^{\frac{3d}{2}+1} \pi^{\frac{d}{2}}C^{-1/\lambda}}{d\Gamma(\frac{d}{2}) F(1)}
\int_0^\infty G(s)s^{-1-\frac{d}{2\lambda}}\ds$. 
\label{lambda}
\end{corollary}
\proof
$\Psi(u) \geq C u^\lambda$ gives $\bern^{-1}(u)\leq C^{-1/\lambda} u^{1/\lambda}$. Then the corollary
follows.
\qed

In some  special cases of Bernstein functions $\bern$ we can derive more explicit forms of the
Lieb-Thirring inequality.

\section{Specific cases}

\subsection{Fractional Schr\"odinger operators (symmetric $\alpha$-stable \\ processes)}
Let $\Psi(u) = (2u)^{\alpha/2}$ and $H^\bern = (-\Delta)^{\alpha/2}$. Throughout this section we
suppose that $0<\alpha \leqslant 2$. Define the quadratic form
\eq{p3}
Q(f, g)= ((-\Delta)^{\alpha/4}f,(-\Delta)^{\alpha/4}g) -(|V|^\han f, |V|^\han g).
\en
Boundedness from below of the cases $\alpha=1$ and $\alpha = 2$ is proven in \cite{lilo01}.
\bl{bound}
Let $V\in L^{d/\alpha}(\BR)+L^\infty(\BR)$.
Then $V$ is form bounded with respect to $(-\Delta)^{\alpha/2}$ with a relative bound strictly
smaller than 1. In particular, we have that 
$\d \inf_{f\in D((-\Delta)^{\alpha/4})} Q(f,f)>-\infty$.
\el
\proof
Let $I_\alpha=(-\Delta)^{-\alpha/2}$ be the operator of the Riesz potential. Recall the Sobolev
inequality $\|I_\alpha f\|_q\leqslant C\|f\|_p$ for $\d q=\frac{pd}{d-\alpha p}$ and $d>\alpha p$.
From this we obtain
\eq{reb}
\|f\|_q\leqslant C\|(-\Delta)^{\alpha/2} f\|_p
\en
with some constant $C$. Hence it follows that
\eq{relativebound}
\|(-\Delta)^{\alpha/4}f\|_2^2\geqslant \frac{1}{C} \|f\|_{\frac{2d}{d-2\alpha}}^2
\geqslant \frac{1}{C}(|V|^\han f, |V|^\han f)\|V\|^{-1}_{d/\alpha}.
\en
The estimate gives $Q(f,f)\geqslant 0$ when $\|V\|_{d/\alpha}<1/C$. Let $V(x)=v(x)+w(x)$ be such that
$v\in L^{d/\alpha}(\BR)$ and $w\in L^\infty(\BR)$. Then there is a bounded function $\lambda(x)$
such that $h=v-\lambda$ satisfies that $\|h\|_{d/\alpha}<1/C$. Thus $V=h+(w+\lambda)$ and $w+\la\in
L^\infty(\BR)$, and the lemma follows.
\qed

\bc{l2}
Let $\bern(u)=(2u)^{\alpha/2}$ and let Assumption \ref{v} hold. If  $V\in L^{d/\alpha}(\BR)$,
then there exists a constant ${\bf L}_{\AP, d}$ independent of $V$ such that
\eq{lt1}
N_0(V) \leqslant {\bf L}_{\AP, d} \int_{\RR^d} |V(x)|^{d/\alpha} \dx ,\quad 0<\alpha\leqslant 2.
\en
\ec
\proof
We have that
\eq{rs01}
{\rm P}^\bern_t(0)=\frac{1}{(2\pi)^d} \int_{\RR^d} e^{-t|\xi|^\alpha}\dxi  = \frac{C(\AP, d)}{t^{d/\AP}},
\en
where $C(\AP, d) = \frac{\s\!\lk {\rm S}_{d-1}\rk \GA\lk d/\AP \rk}{\AP(2\pi)^d}$. Thus the corollary
follows from Theorem \ref{main} with the constant prefactor
\[
{\bf L}_{\AP, d} = \frac{C(\AP, d)}{F(1)}\int_{0}^{\infty} s^{-1-d/\AP}G(s)\ds.
\]
\qed
\noindent
This proof was obtained by hand through direct heat kernel estimates, however, the result also follows
by either of Corollaries  \ref{scale} or \ref{lambda}.

\subsection{Relativistic Schr\"odinger operators (relativistic Cauchy \\ processes)}
Let $\bern(u)=\sqrt{2u+m^2}-m$ and $H^\bern = (-\Delta + m^2)^{1/2}-m$. By using \kak{relativebound} we
derive that
\eq{reltiveboundrelativisticcase}
\|(-\Delta+m^2)^{1/4}f\|_2^2\geqslant \|(-\Delta)^{1/4}f\|_2^2 \geqslant
\frac{1}{C} \|f\|_{\frac{2d}{d-2}}^2 \geqslant
\frac{1}{C}(|V|^\han f, |V|^\han f)\|V\|^{-1}_{d}.
\en
Hence $V\in L^{d/2}(\BR)$ is relatively form bounded with respect to $(-\Delta+m^2)^{1/2}-m$ with relative bound
strictly smaller than 1.
\bc{l3}
Let $\bern(u)=\sqrt{2u+m^2}-m$.
Let Assumption \ref{v} hold, and suppose that $V\in L^d(\RR^d)$ if $m=0$, and $V\in L^{d/2}(\RR^d)\cap L^d(\RR^d)$
if $m\not=0$. Then there exist ${\bf L}_{1,d}^{(1)}, {\bf L}_{1,d}^{(2)}$ and ${\bf L}_{1, d}^{(3)}$ independent
of $V$ such that
\eq{bs18}
\begin{array}{ll}
\d N_{0}(V)\leqslant {\bf L}_{1, d}^{(1)} \int_{\RR^d}|V(x)|^{d}\dx &m=0\\ \\
\d N_{0}(V)\leqslant {\bf L}_{1, d}^{(2)} \int_{\RR^d} |V(x)|^{d}\dx + {\bf L}_{1, d}^{(3)}\int_{\RR^d} |V(x)|^{d/2}\dx
&m\not=0.
\end{array}
\en
\ec
\proof
The proof for $m=0$ can be reduced to Corollary \ref{l2} with $\AP=1$. Let $m>0$. We  have
\[
{\rm P}^\bern_t(0)= \frac{1} {(2\pi)^d}\int_\BR  e^{-t(\sqrt{|\xi|^2+m^2}-m)}\dxi  .
\]
A computation (see Corollary \ref{l3-1} below) gives
\eq{c1}
{\rm P}^\bern_{t}(0)\leqslant \frac{C_1(d)}{t^d}+\frac{C_2(d)}{t^{d/2}}
\en
with some positive constants $C_1(d)$, and $C_2(d)$. Hence we have
$$
N_0(V) \leqslant \frac{1}{F(1)} \left( C_1(d) \int_{\BR} \dx \int_0^\infty \frac{\ds }{s^{1+d}} G(s) |V(x)|^d
+ C_2(d)  \int_{\BR} \dx \int_0^\infty \frac{\ds }{s^{1+{d/2}}} G(s) |V(x)|^{d/2} \right)
$$
for $m\not=0$. Thus the corollary follows with
\begin{eqnarray*}
&&
{\bf L}_{1, d}^{(1)} = \frac{2(d-1)!}{(4\pi)^{d/2}\Gamma(d/2)} \frac{1}{F(1)} \int_0^\infty s^{-1-d}G(s)\ds\\
&&
{\bf L}_{1, d}^{(2)} = \frac{2^{3d/2} (d-1)!}{F(1)} \int_0^\infty s^{-1-d}G(s)\ds  \\
&&
{\bf L}_{1, d}^{(3)} = \frac{2^{-1+3d/4}m^{d/2}\Gamma(d/2)}{F(1)} \int_0^\infty s^{-1-\frac{d}{2}}G(s)\ds.
\end{eqnarray*}
\qed

\subsection{Fractional relativistic Schr\"odinger operators (relativistic $\alpha$-stable processes)}
Let $\bern(u)=({2u+m^{2/\AP}})^{\AP/2}-m$ and $H^\bern = ({-\Delta+m^{2/\AP}})^{\AP/2}-m$. Using
\kak{relativebound} we can also derive that
\eq{reltiveboundrelativisticcase2}
\|(-\Delta+m^{2/\AP})^{\AP/4}f\|_2^2
\geqslant
\|(-\Delta)^{\AP/4}f\|_2^2\geqslant \frac{1}{C} \|f\|_{\frac{2d}{d-2\alpha}}^2
\geqslant \frac{1}{C}(|V|^\han f, |V|^\han f)\|V\|^{-1}_{d/\alpha}.
\en
Hence $V\in L^{d/\AP}(\BR)$ is relatively form bounded with respect to $(-\Delta+m^{2/\AP})^{\AP/2}-m$
with relative bound strictly smaller than 1.
\bc{l3-1}
Let $\bern(u)=(2u+m^{2/\AP})^{\AP/2}-m$, $\alpha\not=1,2$.
Let Assumption \ref{v} hold, and suppose that $V\in L^{d/\AP}(\RR^d)$ if $m=0$,
and $V\in L^{d/\AP}(\RR^d)\cap L^{d/2}(\BR)$
if $m\not=0$. Then there exist ${\bf L}^{(1)}_{\AP, d}, {\bf L}_{\AP, d}^{(2)}$ and ${\bf L}_{\AP, d}^{(3)}$,
independent of $V$ such that
\eq{bs18-1}
\begin{array}{ll}
\d N_{0}(V)\leqslant {\bf L}_{\AP, d}^{(1)} \int_{\RR^d}|V(x)|^{d/\AP}\dx &m=0\\ \\
\d N_{0}(V)\leqslant {\bf L}_{\AP, d}^{(2)} \int_{\RR^d} |V(x)|^{d/\AP}\dx +
{\bf L}_{\AP, d}^{(3)}\int_{\RR^d} |V(x)|^{d/2}\dx &m\not=0.
\end{array}
\en
\ec
\proof
For $m=0$, we adopt the proof of Corollary \ref{l2}. Let $m>0$, then
\eq{frs01}
{\rm P}^\bern_t(0)= \frac{\s\!\lk{\rm S}_{d-1}\rk}{(2\pi)^d}\int_0^\infty
e^{-t((r^2+m^{2/\AP})^{\AP/2}-m)} r^{d-1}\dr.
\en
Using the inequality $u^{\AP/2}-1 \leqslant \frac{\AP}{2}(u-1)$, $0\leqslant u \leqslant 1$, for $\AP \in (0,2)$, and the
substitution $u = m^{2/\AP}/(r^2 + m^{2/\AP})$ it follows that
\eq{frs02}
\lk r^2 + m^{2/\AP} \rk^{\AP/2} - m \geqslant \frac{\AP}{2}r^2 \lk r^2 + m^{2/\AP} \rk^{(\AP/2)-1}.
\en
Assuming that $r \leqslant m^{1/\AP}$, i.e., $r^2 + m^{2/\AP} \leqslant 2 m^{2/\AP}$, it follows from (\ref{frs02}) that
\eq{frs03}
\lk r^2 + m^{2/\AP} \rk^{\AP/2} - m \geqslant \frac{\AP}{2} \frac{ r^2}{\lk 2m^{2/\AP} \rk^{1-\AP/2}}.
\en
If $r > m^{1/\AP}$, i.e., $2r^2 > r^2 + m^{2/\AP}$, then it follows that
\eq{frs04}
\lk r^2 + m^{2/\AP} \rk^{\AP/2} - m \geqslant \frac{\AP}{2^{2-\AP/2}} r^\AP.
\en
Therefore, using (\ref{frs03}) and (\ref{frs04}) in (\ref{frs01}), write
\[
\int_0^\infty e^{-t((r^2+m^{2/\AP})^{\AP/2}-m)} r^{d-1}\dr
\leqslant \int_{r \leqslant m^{1/\AP}} e^{-\frac{\AP r^2}{2\lk 2m^{2/\AP} \rk^{1-\AP/2}}t} r^{d-1}\dr +
\int_{r > m^{1/\AP}} e^{-\frac{\AP r^\AP}{2^{2-\AP/2}}t} r^{d-1}\dr.
\]
For the first integral, set $u = \frac{\AP r^2}{2\lk 2m^{2/\AP} \rk^{1-\AP/2}}t$ to obtain
\eq{frs05}
\int_{r \leqslant m^{1/\AP}} e^{-\frac{\AP r^2}{2\lk 2m^{2/\AP} \rk^{1-\AP/2}}t} r^{d-1}\dr \leqslant
\frac{K_{1}^{d/2}}{2 t^{d/2}} \int_{0}^{\infty} e^{-u} u^{(d/2)-1}\du = \frac{C_{2}(\AP,d)}{t^{d/2}},
\en
where $C_{2}(\AP,d) = \frac{K_{1}^{d/2}\GA\lk d/2 \rk}{2}$, and $K_{1} = \frac{2}{\AP}\lk 2m^{2/\AP} \rk^{1-\AP/2}$.
For the second integral similarly we obtain that
\eq{frs06}
\int_{r > m^{1/\AP}} e^{-\frac{\AP r^\AP}{2^{2-\AP/2}}t} r^{d-1}\dr \leqslant
\frac{1}{\AP}\frac{K_{2}^{d}}{t^{d/\AP}} \int_{0}^{\infty} e^{-u} u^{(d/\AP)-1}\du = \frac{C_{3}(\AP,d)}{t^{d/\AP}},
\en
where $C_{3}(\AP, d) = \frac{K_{2}^{d}\GA\lk d/\AP \rk}{\AP}$, and $K_{2} = \lk \frac{2^{2-\AP/2}}{\AP}\rk^{1/\AP}$.
Thus, using the results of (\ref{main18}) and (\ref{frs01}) together with (\ref{frs05}) and (\ref{frs06}), we find
the positive constants
\begin{align*}
&{\bf L}_{\AP, d}^{(2)} = \frac{C_{2}(\AP, d)}{F(1)}\int_{0}^{\infty} s^{-1-d/2}G(s)\ds,\\
&
{\bf L}_{\AP, d}^{(3)} = \frac{C_{3}(\AP, d)}{F(1)}\int_{0}^{\infty} s^{-1-d/\AP}G(s)\ds
\end{align*}
such that (\ref{bs18-1}) holds for $m \not= 0$. Thus the corollary follows.
\qed

\subsection{Sums of different stable generators}
Let $\bern(u)=(2u)^{\alpha/2}+(2u)^{\beta/2}$, $0 < \alpha, \beta < 2$, $\alpha \neq \beta$, and
$H^\bern=(-\Delta)^{\alpha/2} +(-\Delta)^{\beta/2}+V$,
acting in  $\LR$. Relative boundedness of $V$ follows similarly as in Lemma \ref{bound}, whenever
$V\in L^{d/\alpha}(\BR)\cap L^{d/\beta}(\BR)$. This is an example in which Corollary \ref{scale} does
not apply, however, we have the following result.
\bc{2stableprocesses}
Suppose that Assumption \ref{v} holds and $V\in L^{d/\alpha}(\BR)\cap L^{d/\beta}(\BR)$. Then
\begin{align}
N_0(V)\leq {\bf L}_{\alpha} \int_\BR|V(x)|^{d/\alpha}\dx +  {\bf L}_{\beta} \int_\BR|V(x)|^{d/\beta}\dx,
\end{align}
where
$$
{\bf L}_{\alpha}=\frac{c}{F(1)}\int _0^\infty s^{-1-d/\alpha}G(s) \ds, \quad
{\bf L}_{\beta}=\frac{c}{F(1)}\int _0^\infty s^{-1-d/\beta}G(s) \ds.
$$
\ec
\proof
It is known \cite{CK08} that
\begin{align}
P_t^\bern(0)\leq c \left(t^{-\frac{d}{\alpha}} \wedge t^{-\frac{d}{\beta}}\right), \quad t > 0
\end{align}
with some constant $c>0$. Then by \kak{main18} we obtain the claim.
\qed

\subsection{Jump-diffusion operators}
Let $\bern(u)= u + b u^{\AP/2}$, $\AP \in (0,2)$, and $b \in (0, 1]$. Then we have
$H^\bern =-\Delta+b(-\Delta)^{\AP/2}+V$.
By \kak{relativebound} we see that when $V\in L^{d/\AP}(\BR)\cup L^{d/2}(\BR)$, $V$ is relatively
form bounded with respect to $-\Delta+b (-\Delta)^{\AP/2}$ with relative bound strictly smaller than 1.
\bc{l5-1}
If Assumption \ref{v} holds and $V \in L^{\frac{d}{2}+\frac{d}{\AP}}(\BR)$, then
\eq{bs21}
N_0(V)\leq {\bf L} \int_\BR|V(x)|^{d/2}\dx + {\bf L}_{\alpha} \int_\BR|V(x)|^{d/\alpha}\dx,
\en
where
$$
{\bf L}=\frac{c}{F(1)}\int _0^\infty s^{-1-d/2}G(s) \ds, \quad
{\bf L}_{\alpha}=\frac{c}{F(1)}\int _0^\infty s^{-1-d/\alpha}G(s) \ds.
$$
\ec
\proof
In this case it is known \cite{cks11} that with some $c>0$
\begin{align*}
p_{t}^{b}(x-y) \leq \lk t^{-d/2} \wedge (bt)^{-d/\AP}\rk \wedge \lk t^{-d/2}e^{-|x-y|^{2}/ct} +
(bt)^{-d/\AP} \wedge \frac{bt}{|x-y|^{d+\AP}} \rk,
\end{align*}
and in the same way as in the previous examples the result follows.
\qed

\section*{Acknowledgments}
It is a pleasure to thank Zoran Vondra\v{c}ek for pointing out reference \cite{JKLS} to us.
FH acknowledges support of Grant-in-Aid for Science Research (B) 20340032  and Grant-in-Aid for
Challenging Exploratory Research 22654018 from JSPS. FH also thanks the hospitality of Universit\'e
Paris Sud at Orsay, where part of this work was done. JL thanks ICMS Edinburgh for a RiG grant
sponsoring the workshop ``Functional Integration Methods for Non-Local Operators" (2011), and IHES
Bures-sur-Yvette for a visiting fellowship.

\end{document}